\documentclass[review,11pt]{elsarticle} 
\usepackage{algorithm}
\usepackage{algpseudocode}
\usepackage{graphicx}
\usepackage{comment}
\usepackage{amsmath}
\usepackage{subfigure}

\usepackage{booktabs} 
\usepackage{color}
\usepackage[english]{babel}
\usepackage{picinpar}
\usepackage{graphicx}
\usepackage{moresize}
\usepackage{amssymb}
\usepackage{url}
\usepackage{subfigure}
\usepackage{amsmath}

\usepackage{algorithm}

\begin{document}
\begin{frontmatter}

\title{An Analysis of How COVID-19 Shaped the Realm of Online Gaming and Lesson Delivery}
\author{Yingwei Cheng and Nicholas Milikich}
\address{Department of Computer Science and Engineering\\
University of Notre Dame\\
Notre Dame, IN 46556}

\end{frontmatter}
        \section*{Abstract}\label{sec:abstract}
The COVID-19 pandemic has forced schools and universities to adapt to remote learning, and online gaming has emerged as a tool for education. Educational games can make learning fun and engaging, help students develop important skills like problem-solving and collaboration, and reach students who are struggling with traditional learning methods. While there are concerns about the potential drawbacks of online gaming in education, its benefits are clear. As the pandemic continues to disrupt education, online gaming is likely to become an increasingly important tool for teachers and students alike.

	\section{Introduction} \label{sec:context}
The COVID-19 pandemic has changed the way we live our lives in many ways, including how we approach education. With many schools and universities closed or offering remote learning, students, and teachers have had to adapt to new ways of teaching and learning. One unexpected solution to this problem has been the rise of online gaming as a tool for education~\cite{gandolfi2022updating}.

Online gaming has been around for decades, but during the pandemic, it has seen a surge in popularity. With people stuck at home and looking for ways to stay entertained, online gaming has become a way to connect with friends and family, as well as a way to explore new worlds and learn new skills~\cite{aguirre2021gender}. The gaming industry has responded to this demand by creating educational games that can help students of all ages learn and develop important skills~\cite{ding2020challenges}.

One of the main benefits of online gaming in education is that it can make learning fun. Many students struggle to stay engaged in traditional learning environments, but online games can capture their attention and motivate them to keep learning. Educational games can cover a wide range of topics, from math and science to history and literature. By incorporating game mechanics like achievements and rewards, these games can help students stay motivated and engaged in their learning.

Another benefit of online gaming in education is that it can help students develop important skills like problem-solving, critical thinking, and collaboration. Many online games require players to work together to solve challenges or overcome obstacles, which can help students develop teamwork skills. In addition, many educational games require players to think critically and solve puzzles, which can help them develop problem-solving skills~\cite{shihab2018determinants,ravichandran2023classification}.

Online gaming can also be a way to reach students who are struggling with traditional learning methods. For example, students with learning disabilities or attention deficit hyperactivity disorder (ADHD) may find it easier to learn through interactive games that allow them to move at their own pace and receive immediate feedback~\cite{tzagkaraki2021exploring}. Online games can also be tailored to the individual needs of each student, allowing them to focus on areas where they need extra help~\cite{ding2020challenges}.

Of course, there are also some potential drawbacks to using online gaming in education. One concern is that students may become too focused on gaming and neglect other important aspects of their education. Another concern is that not all online games are created equal, and some may not be effective at teaching the skills they claim to promote~\cite{rashid2015espionage}.

Despite these concerns, the benefits of online gaming in education during the pandemic are clear. Online games can make learning more fun and engaging, help students develop important skills, and reach students who may be struggling with traditional learning methods. As the pandemic continues to disrupt education, it is likely that online gaming will become an increasingly important tool for teachers and students alike~\cite{shihab2017sensing}.

\begin{figure}[!htb]
    \centering
    \includegraphics[width=13cm]{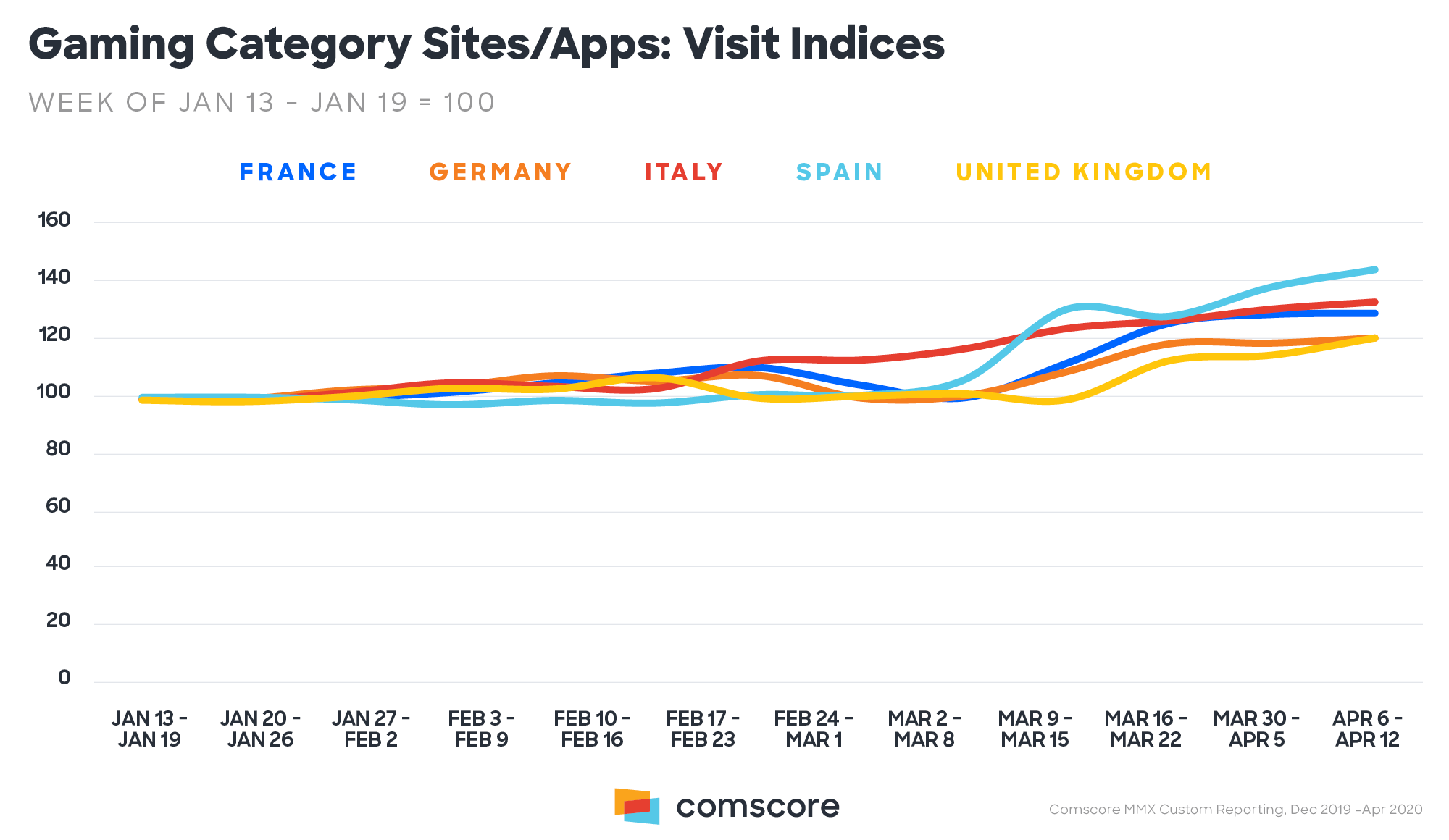}
    \vspace{-0.2in}
    \caption{Patterns of Online Gaming During the Pandemic}
    \vspace{-0.1in}
    \label{fig:bali}
\end{figure}

	\section{Challenges}\label{sec:chhallenge}
While online gaming has emerged as a tool for education during the pandemic, there are several challenges that need to be addressed.

Firstly, there is a concern that students may become too immersed in online gaming and neglect other important aspects of their education. It is crucial to strike a balance between gaming and other forms of learning to ensure that students do not fall behind in their studies~\cite{hew2023online}.

Secondly, not all online games are created equal, and some may not be effective at teaching the skills they claim to promote. It is important to evaluate educational games rigorously and ensure that they align with learning objectives~\cite{silva2021dilemmas}.

Thirdly, not all students have access to the technology needed to participate in online gaming. This digital divide can prevent some students from accessing educational opportunities, and it is important to address this gap to ensure that all students have equal access to learning opportunities~\cite{georgiou2021bridging}.

Fourthly, online gaming requires a level of self-discipline, and some students may struggle to stay motivated without the structure and routine of traditional learning environments. Teachers may need to provide additional support and guidance to ensure that students stay on track and engaged in their learning~\cite{muksin2021level}.

Finally, online gaming can pose risks to student privacy and security~\cite{asgari2021observational}. Schools and universities need to ensure that appropriate measures are in place to protect students' personal information and prevent unauthorized access to educational materials~\cite{crepax2022information}.

Addressing these challenges will be crucial for leveraging the potential of online gaming as a tool for education during the pandemic and beyond.
	\section{Future Directions} \label{sec:prop}
As the use of online gaming in education continues to grow during the pandemic, there is a need for further research to better understand its potential and limitations. Some areas for future research include:

Effectiveness of different types of educational games: There is a need to evaluate the effectiveness of different types of educational games in teaching specific skills and subjects. This research can help identify the most effective games and inform the development of new educational games.

Impact on student learning outcomes: There is a need to assess the impact of online gaming on student learning outcomes, including academic achievement, skill development, and motivation. This research can help determine whether online gaming can complement or replace traditional learning methods.

Role of teacher and student factors: There is a need to investigate the role of teacher and student factors in the effective use of online gaming in education. This research can help identify best practices for integrating online gaming into teaching and learning and provide insights into the characteristics of successful learners.

Digital divide and equity issues: There is a need to address issues of equity and access related to the digital divide~\cite{zhang2020transres}. Research can help identify strategies for bridging the gap and ensuring that all students have equal access to online gaming and other forms of educational technology~\cite{hasan2016development}.

Student privacy and security: There is a need to investigate the privacy and security implications of using online gaming in education. Research can help identify best practices for protecting student data and ensuring that online gaming is used safely and responsibly.

By addressing these areas of research, we can better understand the potential of online gaming as a tool for education during the pandemic and beyond, and develop effective strategies for using it to enhance student learning and engagement.

\begin{figure}[!htb]
    \centering
    \includegraphics[width=12cm]{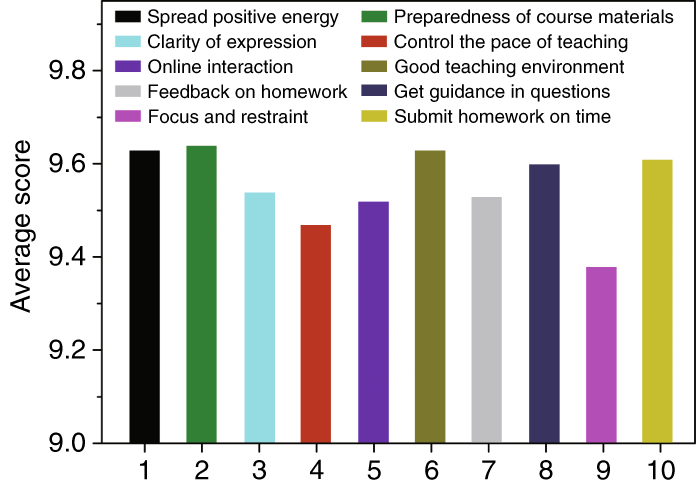}
    \vspace{-0.2in}
    \caption{Impact of Online Education Online on the Education Component}
    \vspace{-0.1in}
    \label{fig:gianyar}
\end{figure}
	\section{Conclusion}
In conclusion, online gaming has emerged as a tool for education during the COVID-19 pandemic. As schools and universities have shifted to remote learning, educational games have become a way to engage students and teach important skills. While there are challenges to using online gaming in education, such as concerns about student engagement and equity issues, the benefits are clear. Online gaming can make learning fun and engaging, help students develop important skills, and reach students who are struggling with traditional learning methods. Looking to the future, further research is needed to better understand the potential of online gaming in education and to address the challenges that arise. By evaluating the effectiveness of different types of educational games, assessing the impact of online gaming on student learning outcomes, investigating the role of teacher and student factors, addressing issues of equity and access related to the digital divide, and ensuring student privacy and security, we can continue to leverage the potential of online gaming as a tool for education during the pandemic and beyond.
\bibliographystyle{elsarticle-num}
\bibliography{ref.bib}

\begin{thebibliography}{10}
\expandafter\ifx\csname url\endcsname\relax
  \def\url#1{\texttt{#1}}\fi
\expandafter\ifx\csname urlprefix\endcsname\relax\def\urlprefix{URL }\fi
\expandafter\ifx\csname href\endcsname\relax
  \def\href#1#2{#2} \def\path#1{#1}\fi

\bibitem{gandolfi2022updating}
E.~Gandolfi, E.~Vongunten, R.~Shihab, Updating the game for a new today:
  In-service teachers’ perspectives on online gaming and education during the
  pandemic, in: Society for Information Technology \& Teacher Education
  International Conference, Association for the Advancement of Computing in
  Education (AACE), 2022, pp. 1648--1653 (2022).

\bibitem{aguirre2021gender}
C.~Aguirre, K.~Harrigian, M.~Dredze, Gender and racial fairness in depression
  research using social media, arXiv preprint arXiv:2103.10550 (2021).

\bibitem{ding2020challenges}
K.~Ding, K.~Shu, Y.~Li, A.~Bhattacharjee, H.~Liu, Challenges in combating
  covid-19 infodemic--data, tools, and ethics, arXiv preprint arXiv:2005.13691
  (2020).

\bibitem{shihab2018determinants}
S.~R. Shihab, Determinants of high enrolment and school dropouts in primary and
  lower secondary schools: A comparative educational appraisal among south
  asian countries, IOSR Journal of Humanities and Social Science 23~(5) (2018)
  72--81 (2018).

\bibitem{ravichandran2023classification}
B.~D. Ravichandran, P.~Keikhosrokiani, Classification of covid-19
  misinformation on social media based on neuro-fuzzy and neural network: A
  systematic review, Neural Computing and Applications 35~(1) (2023) 699--717
  (2023).

\bibitem{tzagkaraki2021exploring}
E.~Tzagkaraki, S.~Papadakis, M.~Kalogiannakis, Exploring the use of educational
  robotics in primary school and its possible place in the curricula, in:
  Education in \& with Robotics to Foster 21st-Century Skills: Proceedings of
  EDUROBOTICS 2020, Springer, 2021, pp. 216--229 (2021).

\bibitem{rashid2015espionage}
M.~T. Rashid, P.~Chowdhury, M.~K. Rhaman, Espionage: A voice guided
  surveillance robot with dtmf control and web based control, in: 2015 18th
  International Conference on Computer and Information Technology (ICCIT),
  IEEE, 2015, pp. 419--422 (2015).

\bibitem{shihab2017sensing}
S.~R. Shihab, N.~Sultana, Sensing the necessity and impacts of private tuition
  in english among secondary students in khulna, bangladesh, Global Journal of
  Human-Social Science Research 17~(6) (2017).

\bibitem{hew2023online}
J.-J. Hew, V.-H. Lee, S.-T. T’ng, G.~W.-H. Tan, K.-B. Ooi, Y.~K. Dwivedi, Are
  online mobile gamers really happy? on the suppressor role of online game
  addiction, Information Systems Frontiers (2023) 1--33 (2023).

\bibitem{silva2021dilemmas}
A.~J. F.~d. Silva, C.~C.~d. Silva, R.~d.~G. Tin{\^o}co, A.~C.~d. Ara{\'u}jo,
  L.~Ven{\^a}ncio, L.~Sanches~Neto, E.~d.~S. Freire, W.~Lazaretti~da
  Concei{\c{c}}{\~a}o, Dilemmas, challenges and strategies of physical
  education teachers-researchers to combat covid-19 (sars-cov-2) in brazil, in:
  Frontiers in Education, Vol.~6, Frontiers Media SA, 2021, p. 583952 (2021).

\bibitem{georgiou2021bridging}
Y.~Georgiou, E.~A. Kyza, Bridging narrative and locality in mobile-based
  augmented reality educational activities: Effects of semantic coupling on
  students’ immersion and learning gains, International Journal of
  Human-Computer Studies 145 (2021) 102546 (2021).

\bibitem{muksin2021level}
S.~N.~B. Muksin, M.~B. Makhsin, A level of student self-discipline in
  e-learning during pandemic covid-19, Procedia of Social Sciences and
  Humanities 1 (2021) 278--283 (2021).

\bibitem{asgari2021observational}
S.~Asgari, J.~Trajkovic, M.~Rahmani, W.~Zhang, R.~C. Lo, A.~Sciortino, An
  observational study of engineering online education during the covid-19
  pandemic, Plos one 16~(4) (2021) e0250041 (2021).

\bibitem{crepax2022information}
T.~Crepax, V.~Munt{\'e}s-Mulero, J.~Martinez, A.~Ruiz, Information technologies
  exposing children to privacy risks: Domains and children-specific technical
  controls, Computer Standards \& Interfaces 82 (2022) 103624 (2022).

\bibitem{zhang2020transres}
Y.~Zhang, R.~Zong, J.~Han, D.~Zhang, T.~Rashid, D.~Wang, Transres: a deep
  transfer learning approach to migratable image super-resolution in remote
  urban sensing, in: 2020 17th annual IEEE international conference on sensing,
  communication, and networking (SECON), IEEE, 2020, pp. 1--9 (2020).

\bibitem{hasan2016development}
S.~M. Hasan, M.~T. Rashid, M.~S.~S. Chowdhury, M.~K. Rhaman, Development of a
  credible and integrated electronic voting machine based on contactless ic
  cards, biometrie fingerprint credentials and pos printer, in: 2016 IEEE
  Canadian Conference on Electrical and Computer Engineering (CCECE), IEEE,
  2016, pp. 1--5 (2016).

\end{thebibliography}

\end{document}